\documentclass[iop]{emulateapj}
\usepackage{lscape}
\usepackage{rotating}
  \received{}
  \revised{}
  \accepted{}

\shortauthors{Tortora et al.} \shorttitle{DM fraction in lens
galaxies}

\def\eSF{\mbox{$\epsilon_{\rm SF}$}}
\def\Re{\mbox{$R_{\rm eff}$}}
\def\RE{\mbox{$R_{\rm E}$}}

\def\mst{\mbox{$M_{\star}$}}

\def\fdm{\mbox{$f_{\rm DM}$}}
\def\lsim{\mathrel{\rlap{\lower3.5pt\hbox{\hskip0.5pt$\sim$}}
    \raise0.5pt\hbox{$<$}}}                
\def\gsim{~\rlap{$>$}{\lower 1.0ex\hbox{$\sim$}}}

\begin{document}

\title{Central dark matter trends in early-type galaxies from strong lensing, dynamics and stellar populations}

\author{C.~Tortora\altaffilmark{1}, N.R.~Napolitano\altaffilmark{2}, A.J.~Romanowsky\altaffilmark{3}, P. Jetzer\altaffilmark{1}}


\altaffiltext{1}{Universit$\rm \ddot{a}$t Z$\rm \ddot{u}$rich,
Institut f$\rm \ddot{u}$r Theoretische Physik, Winterthurerstrasse
190, CH-8057, Z$\rm \ddot{u}$rich, Switzerland}\email{\texttt
ctortora@physik.uzh.ch}

\altaffiltext{2}{INAF -- Osservatorio Astronomico di Capodimonte,
Salita Moiariello, 16, 80131 - Napoli, Italy}

\altaffiltext{3}{UCO/Lick Observatory, University of California,
Santa Cruz, CA 95064, USA}


\begin{abstract}
We analyze the correlations between central dark matter (DM)
content of early-type galaxies and their sizes and ages, using a
sample of intermediate-redshift ($z\sim 0.2$) gravitational lenses
from the SLACS survey, and by comparing them to a larger sample of
$z\sim0$ galaxies. We decompose the deprojected galaxy masses into
DM and stellar components using combinations of strong lensing,
stellar dynamics, and stellar populations modeling. For a given
stellar mass, we find that for galaxies with larger sizes, the DM
fraction increases and the mean DM density decreases, consistently
with the cuspy halos expected in cosmological formation scenarios.
The DM fraction also decreases with stellar age, which can be
partially explained by the inverse correlation between size and
age. The residual trend may point to systematic dependencies on
formation epoch of halo contraction or stellar initial mass
functions. These results are in agreement with recent findings
based on local galaxies by \cite{NRT10} and suggest negligible
evidence of galaxy evolution over the last $\sim 2.5$~Gyr other
than passive stellar aging.
\end{abstract}


\section{Introduction}\label{sec:intro}

Early-type galaxies (ETGs) are among the most massive systems in
the Universe. They are on average metal-rich, dust and gas poor,
and formed their stars in early, rapid events (e.g.,
\citealt{Thomas2005,Fontanot+09}). The lack of easily
interpretable dynamical tracers, such as the cold gas in spiral
galaxies, has made the mass mapping of these systems very
difficult in the outer regions where dark matter (DM) is expected
to be dominant, although discrete tracers (planetary nebulae and
globular clusters; e.g., \citealt{Napolitano+09,Romanowsky+09})
are providing a clearer view of the DM halos. The mass content of
these galaxies' central regions has on the other hand been
extensively investigated (e.g., \citealt{Gerhard+01};
\citealt{Cappellari+06}; \citealt[T+09 hereafter]{Tortora2009a}),
with building evidence that the central DM fraction (\fdm) is an
increasing function of total stellar mass ($M_\star$), providing
the main driver for the tilt of the fundamental plane (e.g.,
T+09).

New insights into the galaxy assembly process are emerging from
joint analyses of the structural and star formation properties of
nearby ETGs (e.g., \citealt{Gargiulo+09,Graves10}; \citealt[NRT10
hereafter]{NRT10}). A key discovery of NRT10 is an
anti-correlation between \fdm\ and galaxy stellar age, such that
older galaxies (at a fixed $M_\star$) have lower \fdm. In the
context of $\Lambda$CDM halos, this trend can be partially
explained by an anti-correlation between galaxy sizes and ages,
with the remaining effect apparently driven by variations in star
formation efficiency, stellar initial mass function (IMF), or DM
distribution (e.g., adiabatic contraction, AC hereafter). As
discussed in NRT10, such correlations would have deep implications
for the assembly histories of ETGs, but critically need to be
confirmed by independent analyses.

Gravitational lenses offer a unique tool to map the mass profile
in galaxies over a range of redshifts. The database of lenses is
growing quickly thanks to ongoing surveys (e.g., SLACS:
\citealt[A+09 hereafter]{Auger+09}; COSMOS: \citealt{Faure+08}).
In particular, the SLACS survey has collected $\gsim 80$ secure
lenses, which is a sample comparable to the total number of lenses
discovered since the late 70s from other campaigns or by
serendipity (e.g. \citealt{Covone+09}). Here, we will use the
SLACS sample data to extend the analysis of NRT10, using both
lensing and dynamics as independent probes of total mass, and
providing a higher-redshift ($z\sim0.2$) comparison to the
$z\sim0$ galaxies previously studied\footnote{In the paper, we use
a cosmological model with $(\Omega_{m}, \, \Omega_{\Lambda}, \, h)
= (0.3, \, 0.7, \, 0.7)$, where $h = H_{0}/100 \, \textrm{km} \,
\textrm{s}^{-1} \, \textrm{Mpc}^{-1}$ (\citealt{WMAP2}),
corresponding to a universe age today of $t_{\rm univ}=13.5 \, \rm
Gyr$.}.

\section{Data sample and analysis}\label{sec:sample}

\subsection{Galaxy Samples}

The lensing galaxy sample is taken from the SLACS survey
(A+09)\footnote{See also {\tt http://www.slacs.org/}}, which has
been extensively analyzed in other works (e.g.,
\citealt{SLACS1,SLACS5,Gavazzi07,Cardone+09,Treu+09,Grillo+09,CT10}).
The lens galaxy redshift ($z_{l}$) range is $0.05\leq z_l\leq
0.5$, with a median of $z_{l} \sim 0.2$. Our sample is selected:
1) to have a measured Einstein radius \RE\footnote{The Einstein
radii are derived by fitting a singular isothermal ellipsoid (SIE)
profile and are quoted adopting an intermediate-axis
normalization. Five of the galaxies without a measured \RE\ have a
nearby companion while, for other systems, the $HST$ data do not
have sufficient sensitivity to adequately perform the lensing
model.} in Table 3 of A+09, 2) to be classified as elliptical or
S0, 3) to have a measured $V$-band effective radius (measured at
the intermediate axis). Of the 85 lenses from this latest SLACS
release, 66 passed our selection criteria.

As a $z\sim0$ comparison sample, we use the collection of 330 ETGs
over the same mass range analyzed in T+09 and NRT10.

\subsection{Stellar population analysis}

To estimate stellar mass-to-light ratios ($\Upsilon_\star$) and
star formation histories, we analyze spectral energy distributions
(SEDs) based on broad-band Sloan Digital Sky Survey (SDSS)
photometry (namely, $ugriz$). Our general procedure is to adopt a
set of synthetic spectra from the prescription of \citet{BC03}, a
uniform metallicity $Z$, an age $t$ characterizing the time of
star formation onset, and an exponentially declining star
formation rate with timescale $\tau$. For each galaxy, $Z$, $t$,
and $\tau$ are fitted parameters, with the determined
$\Upsilon_{\star}$ based on a \citet{Kroupa01} IMF; uncertainties
on the estimated parameters have been quantified via Monte Carlo
simulations: further details are provided in T+09 and NRT10 along
with explorations of systematic uncertainties and degeneracies.
The only change here is to shift the spectral responses of the
SDSS filters to correspond to the lens redshifts before convolving
with the model SEDs. We have checked that imposing restrictions on
$Z$ or $\tau$, or adopting the stellar populations results from
A+09 or \citet{Grillo+09}, does not qualitatively affect the
results described below.

\subsection{Total mass and dark matter content}

We derive the deprojected total mass from dynamics and lensing
observables, and separate the DM from the stellar components using
the stellar mass estimates discussed in the previous section. We
adopt the stellar effective radius \Re\ as the fiducial reference
point for mass comparisons. For the lens galaxies, \Re\ is
measured in the $V$-band, corresponding approximately to the
rest-frame $B$-band used for the local galaxies. In both galaxy
samples, the mass constraints are generally based on measurements
at smaller radii ($\sim 0.1 \Re$ and $\sim 0.5 \Re$,
respectively), and therefore some extrapolation is required. For
the total mass distribution we adopt a  singular isothermal sphere
(SIS) with density $\rho(r) = \sigma_{\rm SIS}^{2} / (2 \pi G
r^{2})$ (e.g. \citealt{SLACS3}, \citealt{Gavazzi07},
\citealt{Koopmans+09}), where $\sigma_{\rm SIS}$ is an unknown
normalization to be determined by fitting the observables. For the
stars, we adopt a constant-$\Upsilon_*$ mass profile based on the
\citet{H90} model.

To estimate dynamical masses we have used the SDSS stellar
velocity dispersions $\sigma_{\rm SDSS}$, measured within a
circular aperture of $R_{\rm ap} = 1.5''$. Briefly, we have
adopted the spherical Jeans equation to derive the surface
brightness weighted velocity dispersion $\sigma_{\rm ap,SIS}$
within $R_{\rm ap}$ (see T+09 for further details), to be matched
to $\sigma_{\rm SDSS}$. As discussed in T+09, there is some degree
of systematic uncertainty from assumptions of sphericity and
orbital isotropy, which we can now check in the case of the lenses
by using the independent lensing-based masses (which do have their
own uncertainties from mass-sheet degeneracies).

For the lensing mass estimates, we have used the Einstein radius
\RE, to derive  a model independent measurement of projected mass
($M_{\rm proj}$) within \RE, since $M_{E}=M_{\rm proj}(\RE) = \pi
\RE^{2} \Sigma_{\rm crit}$, where $\Sigma_{\rm crit} = c^{2}
D_{s}/4 \pi G D_{l} D_{ls}$, with $D_{s}$, $D_{l}$ and $D_{ls}$
the observer - source, observer - lens and lens - source comoving
angular diameter distances, respectively. Finally we match the
prediction of the SIS model projected mass, $M_{proj,SIS}$ with
$M_{E}$ to have a further constraint on the only free model
parameter for each galaxy, $\sigma_{\rm SIS}$.

The best fitted $\sigma_{\rm SIS}$ can be derived independently
using either technique to estimate the best 3D deprojected mass
profile which we extrapolate to $r = \Re$ to obtain our reference
mass values. Using this approach, we find that lensing and
dynamics provide consistent results, modulo a $\sim$~10\% (and a
scatter of $\sim 25\%$) higher mass from dynamics, corresponding
to a change of $0.03 \pm 0.10$ in $f_{\rm DM}$. Thus, we adopt a
combination of the constraints for our final masses, by minimizing
with respect to $\sigma_{\rm SIS}$ a combined $\chi^2$ function
including one term for dynamics and one for lensing observables,
given by
\begin{equation}
\chi^{2}= \bigg
(\frac{\sigma_{ap}-\sigma_{SDSS}}{\delta_{d}}\bigg)^{2} + \bigg
(\frac{M_{E}-M_{proj,SIS}}{\delta_{l}}\bigg)^{2},
\end{equation}
where $\delta_{d}$ and $\delta_{l}$ are the uncertainties on
$\sigma_{\rm SDSS}$ and $M_{E}$, respectively \footnote{Note that
$\delta_{d}$ is given in Table 3 of A+09 and ranges from $2\%$ to
$19\%$ (with mean $6\%$), while we have assumed a nominal $5\%$
uncertainty on \RE\ which corresponds to a relative error of
$10\%$ on $M_{E}$. However, the results are qualitatively
unchanged if we would assume $\delta_{d}=\delta_{l}$.}.

In the following we will focus on the central 3D deprojected \fdm\
and the mean DM density within \Re , defined as $\langle \rho_{\rm
DM}\rangle= M_{\rm DM}/(4/3 \pi \Re^3)$ where $M_{\rm DM}=M_{\rm
tot}-M_\star$ at \Re{} is the DM mass.

\subsection{Cosmological models}

As in T+09 and NRT10, to interpret the observational results, we
construct a series of toy mass models based on $\Lambda$CDM
cosmological simulations. For each bin in $M_\star$, we use the
average $\Re$-age relations from the combined lens+local sample,
and parameterize the virial DM mass by a star formation efficiency
$\epsilon_{\rm SF}=M_\star/(\Omega_{\rm bar} M_{\rm tot})$, where
$\Omega_{\rm bar}=0.17$ (\citealt{WMAP2}) is the baryon density
parameter. The halo densities are initially characterized as
\citet{NFW} profiles following an average mass-concentration
relation, adjusted by $(1+z)^{-1}$ for the lens galaxies. A recipe
for AC from baryon settling is then applied \citep{Gnedin+04}. The
toy models for $\eSF = 0.03,0.1,0.3$ are shown in both Figs.
\ref{fig: fig1} and \ref{fig: fig3}.

\section{Results: correlations with size and formation epoch}\label{sec:results}

In order to marginalize any correlations with $M_\star$, we group
the galaxies from both samples into bins of common median mass:
$\log M_*/M_\odot\sim 11.6$, $11.3$ and $10.9$. For the lens
galaxies, the corresponding median redshifts are $z_{\rm
med}=0.28, 0.18, 0.13$. The local galaxies sample extends to even
lower masses ($\log M_*/M_\odot \sim 10.4$) with no lens
counterparts.

Our first result (which we do not show for the sake of space) is
that for the lenses, \fdm\ increases on average with $M_\star$
(see \citealt{Cardone+09} and \citealt{CT10} for details). This
result confirms that as in local galaxies (T+09), \fdm\ is also a
main driver of the fundamental plane tilt at $z\sim 0.2$.

Next, following NRT10, we focus on correlations of the DM metrics
with galaxy size and age. For the latter we adopt the look-back
time to the formation epoch in order to put all the galaxies with
different observed redshifts on a common reference frame.

\begin{figure}[t]
{\hspace{-0.59cm}}\includegraphics[width=0.58\textwidth,clip]{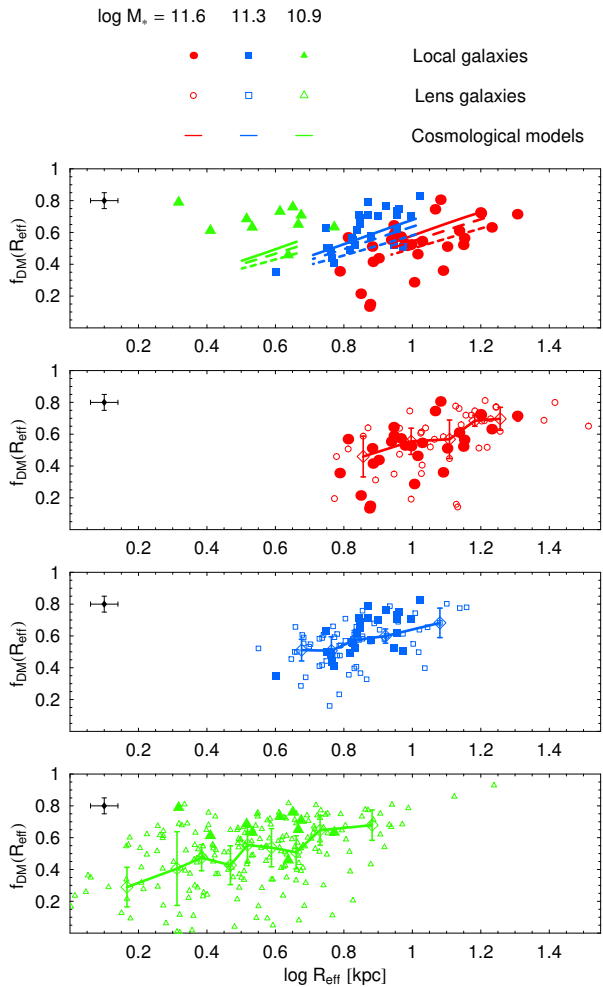}
\caption{Dark matter fraction within an effective radius (\Re) as
a function of \Re. The lens and local galaxies are shown as filled
and open symbols, respectively. For the latter, open symbols with
error bars show the median and $\pm 25\%$ values. Typical
$1\sigma$ uncertainties for individual galaxies are shown to the
left. The differently colored symbols denote different bins of
stellar mass, as labeled in the legend in the top. The second
panel from the top shows the combined bins for the lens galaxies
only and includes toy-model $\Lambda$CDM predictions: solid,
long-dashed, and short-dashed curves show star formation
efficiencies of $\epsilon_{\rm SF}=0.03,0.1,0.3$, respectively.
}\label{fig: fig1}
\end{figure}

Fig. \ref{fig: fig1} demonstrates that there is a strong positive
correlation between \fdm\ and \Re, once the galaxies are divided
into mass bins. This may be understood as a larger \Re\ enclosing
a bigger portion of the DM halo; this ``aperture effect'' appears
to be more dominant than the \fdm\ correlation with $M_\star$. The
local and lens samples appear reasonably similar, although the
lens galaxies in the lowest mass bin are systematically higher,
which is an issue we will discuss below. Both samples are in rough
agreement with our $\Lambda$CDM toy model predictions (top panel).

\begin{figure}[t]
\centering
\includegraphics[width=0.45\textwidth,clip]{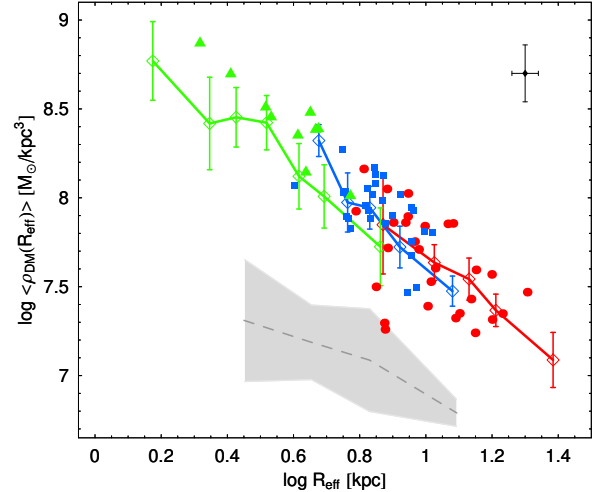}
\caption{Mean DM density within 1 \Re\ as a function of \Re. See
Fig. \ref{fig: fig1} for the meaning of symbols. The grey region
shows an average of spiral galaxy results from
\citet{McGaugh+07}.}\label{fig: fig2}
\end{figure}

Fig.~\ref{fig: fig2} shows that $\langle\rho_{\rm DM}\rangle$
strongly anti-correlates with \Re. Again considering the aperture
effect and assuming DM halo homogeneity, the implication is that
we are measuring a mean DM density profile with radius, with a
best fitted log slope of $\sim -1.7$. As discussed in NRT10, this
steep slope is indicative of cuspy halos, perhaps as induced by
AC. One could suspect that we are getting out what we are putting
in, since our default galaxy model assumes an isothermal total
density profile (with slope $\sim -2$) in order to extrapolate
measurements to $r = \Re$, but we have shown in NRT10 that the use
of an alternative constant-M/L profile yields similar results
(modulo a difference of $0.1-0.2$ in the slope), still fully
consistent with a cuspy contracted halo\footnote{Note that the two
massive galaxies J0157$-$0056 and J0330$-$0020 with $\log \Re \sim
0.9$, which depart from the mean trend, are fitted by a possibly
unrealistic supersolar metallicity; setting $Z$ to the solar
value, the galaxies' densities are incremented by $0.15$ and
$0.25$ dex, respectively.}.

In Fig. \ref{fig: fig2}, we also see that the ETGs have DM
densities substantially larger than those of local spiral
galaxies, which have been suggested to follow a unified halo trend
with dwarf spheroidals \citep{Donato+09,Walker+10b}. This
dichotomy is qualitatively consistent with other findings
(\citealt{Gerhard+01,Thomas+09}; NRT10; \citealt{CT10}), and may
imply different formation mechanisms. The difference might be
assumed as simply caused by ETGs forming from denser late-type
galaxies at earlier epochs, which would yield the corollary
prediction that ETGs with younger stellar ages have less dense
halos because their late-type progenitors were less dense.
However, as we will see below, the opposite trend appears to be
observed.

Finally we consider the \fdm-age dependencies in Fig.~\ref{fig:
fig3}, again using separate $M_\star$ bins. The lens and local
galaxies match up remarkably well in general, showing a clear
trend for lower \fdm\ at older ages. The low-mass lens galaxies
are predominantly young, which is probably a selection effect on
apparent magnitude. The \fdm-age anti-correlation then causes the
overall \fdm\ for this mass bin to be high. In summary, at a fixed
galaxy age and mass, the higher-$z$ sample does not show any
significant difference with the local galaxies. Since we have not
applied any evolutionary corrections, the implication is that in
the last $\sim 2.5$ Gyr (on average) the galaxy populations have
experienced no measurable evolution except stellar
aging\footnote{Here the stellar mass bins can be biased at
different redshifts due to the stellar population evolution and
introduce a spurious tilt in the $\fdm$-age relation. We have
checked from the stellar models that the change in stellar mass
due to stellar evolution might be not larger than $\sim -0.01$ in
$\log \mst$ (due to mass loss), which makes our stellar bins at
different $z$ fairly homogeneous.}.

\begin{figure}[t]
{\hspace{-0.95cm}}\includegraphics[width=0.6\textwidth,clip]{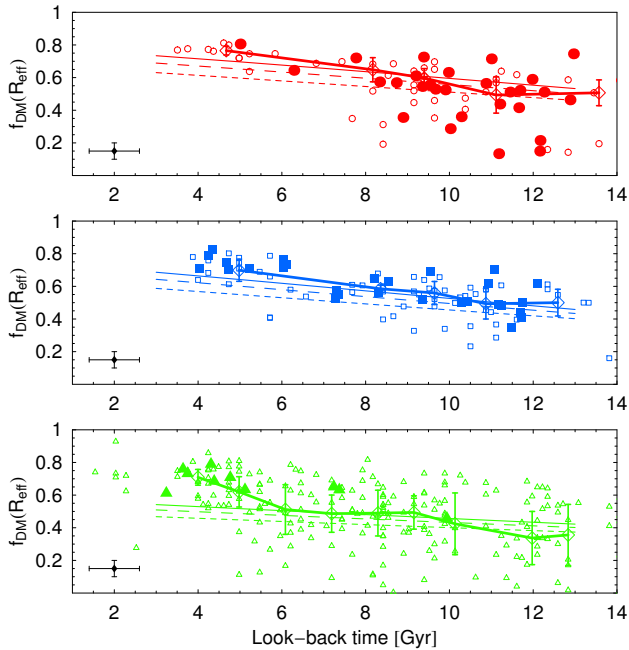}
\caption{DM fraction within 1 \Re\ as a function of galaxy stellar
``age''. See Figs. \ref{fig: fig1} and \ref{fig: fig2} for the
meaning of symbols.}\label{fig: fig3}
\end{figure}

We also include the $\Lambda$CDM toy models in Fig. \ref{fig:
fig3}, and see that some \fdm-age anticorrelation is expected,
which can be traced to the anti-correlation between \Re\ and age.
However, there are indications in every mass bin that the observed
\fdm-age slope is steeper than in the models. Some systematic link
between age and $\epsilon_{\rm SF}$ is possible, but the models
shown in Fig.~\ref{fig: fig3} suggest that this would not be a
strong enough effect, as changes in virial mass do not propagate
strongly to changes in central DM content (and in fact earlier
collapsing halos should have {\it denser} centers, which goes in
the wrong way to explain the observations). The alternatives as
discussed in NRT10 are that AC is more effective in younger
galaxies, or that older galaxies have ``lighter'' IMFs (e.g.,
Kroupa versus \citealt{Salpeter55} for the younger galaxies).

\section{Conclusions}\label{sec:conclusions}

We have analyzed the central DM content of a sample of
intermediate-$z$ lenses from the latest release of the SLACS
survey (A+09). Following the phenomenological framework introduced
in T+09 and NRT10 we have discussed scaling relations between DM
fraction, galaxy size, and formation epoch. Gravitational lensing
and dynamical analyses are used to constrain the total mass
profile, while synthetic spectral populations are used to infer
the stellar mass and other stellar properties such as galaxy age.

ETGs at $z \sim 0.2$ are found to be similar to local ones; future
work will include extending the baseline to higher redshifts. The
somewhat surprising findings of NRT10 are now confirmed with an
independent and arguably more robust data set. The DM fraction
within \Re\ is found to strongly correlate with \Re, because
larger length-scales probe a more DM dominated region. On these
scales, the DM mean density decreases with \Re\ as
$\langle\rho_{\rm DM}\rangle \, \propto \Re^{-1.7}$, which argues
for cuspy DM halos for ETGs out to $z\sim0.5$ (as was discussed in
NRT10 for local galaxies). At a fixed stellar mass and
length-scale, we have found that the DM halos of ETGs are denser
than those of local spiral galaxies, providing a critical test for
the merging formation scenario (see also \citealt{CT10}).

Finally, we have confirmed our earlier finding that central DM
content anti-correlates with stellar age. The strength of this
correlation appears to exceed what is expected from size-age
effects. A fundamental connection between galactic structure and
star formation history is implied, which we propose is a
consequence of variations with formation epoch of either DM halo
contraction or stellar IMF.

In future work, we plan to investigate the impact on these results
of more complex total and DM galaxy profiles (e.g.
\citealt{Cardone05}, \citealt{Tortora2007}) along the lines of
recent work in \cite{Cardone+09} and \cite{CT10}. New high-quality
data are also expected with the advent of future surveys both in
the local Universe and at larger redshifts. Such surveys will
include larger samples of gravitational lenses along with more
detailed spectroscopic information, and could be used to verify
and extend the results presented here, providing a clearer picture
of the physical processes of ETG assembly.


\acknowledgments

We thank the anonymous referee for the suggestions which helped to
improve the paper. CT was supported by the Swiss National Science
Foundation. AJR was supported by National Science Foundation
Grants AST-0808099 and AST-0909237.


\vspace{0.5cm}

{\it Note added in proof.} During the final phase of this
manuscript, there appeared closely related work from \citet[A+10a
and A+10b respectively, hereafter]{Auger+10a, Auger+10b}, using a
combination of strong lensing, and stellar dynamics and
populations, to analyze virtually the same lens-galaxy sample. A
key point of agreement is that A+10a found the strongest
non-trivial correlate with central $f_{\rm DM}$ is \Re, which we
think demonstrates that size rather than mass variation is the
main driver for the fundamental plane tilt.

A+10b also constructed $\Lambda$CDM models, adding constraints
from weak lensing to break the degeneracies between \eSF, IMF, and
halo model (AC or no-AC). Given the same IMF and halo assumptions,
they found SF values that are on average higher than both ours
(e.g. $\eSF \sim 0.3-0.4$ versus $\sim 0.1$ for Salpeter+no-AC),
and typical values found in other studies. Although these
differences may seem large, they involve relatively small changes
in the central DM properties which could be driven by differences
in the modeling methods. We considered three-dimensional DM masses
within \Re, while A+10b analyzed projected masses within $\Re/2$
in combination with large-radius constraints from weak lensing, so
a direct comparison is not straightforward. We hope to track down
the reasons for these discrepant conclusions in future work.

A+10b found that a Salpeter+no-AC model is preferred over a
Chabrier IMF (with or without AC) using direct model fits. Without
weak lensing constraints, our modeling permits either
Salpeter+no-AC or Kroupa+AC solutions (see NRT10 for further
details), while our Figure 2 (or Figure 9 in NRT10) does provide
indirect evidence for very cuspy halos with AC (and a Kroupa IMF).
They also found that the IMF may become heavier with galaxy mass
($\eta > 1$ in their notation). We find that stellar age is the
more important parameter in this context, but if an IMF-mass
relation is demanded, then $\eta < 1$. This difference appears to
be caused at least partially by different stellar populations
models: although we found that varying these models did not alter
the basic age trends, the mass trends are weaker and more
sensitive to the models.


\begin{thebibliography}{}




\bibitem[\protect\citeauthoryear{Auger et al.}{2009}]{Auger+09} Auger M. W., Treu T., Bolton A. S., Gavazzi R., Koopmans L.
V. E., Marshall P. J., Bundy K. \& Moustakas L. A. 2009, ApJ, 705,
1099 (A+09)


\bibitem[\protect\citeauthoryear{Auger et al.}{2010a}]{Auger+10a}  Auger, M. W., Treu, T., Bolton, A. S., Gavazzi, R., Koopmans, L.
V. E., Marshall, P. J., Moustakas, L. A., \& Burles, S. 2010a,
ApJ, submitted (arXiv:1007.2880) (A+10a)


\bibitem[\protect\citeauthoryear{Auger et al.}{2010b}]{Auger+10b} Auger, M. W., Treu, T., Gavazzi, R., Bolton, A. S., Koopmans, L.
V. E., \& Marshall, P. J. 2010b, ApJL, submitted (arXiv:1007.2409)
(A+10b)

\bibitem[\protect\citeauthoryear{Bolton et al. }{2006}]{SLACS1} Bolton, A. S., Burles, S. Koopmans, L. V. E., Treu, T., Moustakas, L.
A. 2006, ApJ, 638, 703

\bibitem[\protect\citeauthoryear{Bolton et al. }{2008}]{SLACS5} Bolton, A. S., Burles, S. Koopmans, L. V. E., Treu, T., et al. 2008a, ApJ, 682,
964

\bibitem[\protect\citeauthoryear{Bruzual \& Charlot }{2003}]{BC03} Bruzual, A. G. \& Charlot, S. 2003, MNRAS, 344,
1000 (BC03)

\bibitem[\protect\citeauthoryear{Cappellari et al.}{2006}]{Cappellari+06}  Cappellari M. et al. 2006, MNRAS, 366, 1126


\bibitem[\protect\citeauthoryear{Cardone et al.}{2005}]{Cardone05} Cardone V. F., Piedipalumbo E. \& Tortora
C. 2005, MNRAS, 358, 1325

\bibitem[\protect\citeauthoryear{Cardone et al. }{2009}]{Cardone+09} Cardone V. F., Tortora C., Molinaro R., Salzano
V. 2009, A\&A, 504, 769


\bibitem[\protect\citeauthoryear{Cardone \& Tortora}{2010}]{CT10} Cardone V. F. \& Tortora C. 2010, MNRAS, submitted (arXiv:1007.3673)


\bibitem[\protect\citeauthoryear{Covone et al.}{2009}]{Covone+09} Covone, G. et al. 2009, ApJ, 691, 531


\bibitem[\protect\citeauthoryear{Donato et al.}{2009}]{Donato+09}  Donato, F. et al. 2009, MNRAS, 397, 1169


\bibitem[\protect\citeauthoryear{Faure et al.}{2008}]{Faure+08} Faure C. et
al. 2008, ApJS, 176, 19


\bibitem[\protect\citeauthoryear{Fontanot et al.}{2009}]{Fontanot+09} Fontanot, F. et al. 2009, arXiv:0901.1130

\bibitem[Gargiulo et al.(2009)]{Gargiulo+09} Gargiulo, A., et al.\ 2009, \mnras, 397, 75

\bibitem[\protect\citeauthoryear{Gavazzi et al. }{2002}]{Gavazzi02} Gavazzi, G., Bonfanti, C., Sanvito, G., Boselli, A., Scodeggio,
M. 2002, ApJ, 576, 135G

\bibitem[\protect\citeauthoryear{Gavazzi et al. }{2007}]{Gavazzi07} Gavazzi, R. et al. 2007, ApJ, 667, 176G


\bibitem[\protect\citeauthoryear{Gerhard et al. }{2001}]{Gerhard+01}  Gerhard, O., Kronawitter, A., Saglia, R. P. \& Bender, R. 2001,
AJ, 121, 1936

\bibitem[\protect\citeauthoryear{Gnedin et al. }{2004}]{Gnedin+04} Gnedin O. Y., Kravtsov A. V., Klypin A. A. \& Nagai D. 2004,
ApJ, 616, 16

\bibitem[Graves \& Faber(2010)]{Graves10} Graves, G.~J., \& Faber, S.~M.\ 2010, ApJ, submitted, arXiv:1005.0014

\bibitem[Grillo et al.(2009)]{Grillo+09} Grillo, C., Gobat, R., Lombardi, M., \& Rosati, P.\ 2009, A\&A, 501, 461

\bibitem[\protect\citeauthoryear{Hernquist }{1990}]{H90}
Hernquist, L. 1990, ApJ, 356, 359

\bibitem[\protect\citeauthoryear{Koopmans et al. }{2006}]{SLACS3} Koopmans, L. V. E., Treu, T., Bolton, A.
S., Burles, S., Moustakas, L. A. 2006, ApJ, 649, 599K

\bibitem[\protect\citeauthoryear{Koopmans et al. }{2009}]{Koopmans+09} Koopmans L. V. E., et al. 2009, ApJ, 703, 51

\bibitem[\protect\citeauthoryear{Kroupa }{2001}]{Kroupa01} Kroupa P., 2001, MNRAS, 322, 231


\bibitem[\protect\citeauthoryear{McGaugh et al.}{2007}]{McGaugh+07} McGaugh, S. S., de Blok,W. J. G., Schombert, J. M., Kuzio de
Naray, R. \& Kim, J. H. 2007, ApJ, 659, 149

\bibitem[\protect\citeauthoryear{Napolitano et al.}{2009}]{Napolitano+09} Napolitano, N.~R., Romanowsky, A.~J., Coccato, L.,
et al. 2009, MNRAS, 393, 329

\bibitem[\protect\citeauthoryear{Napolitano, Romanowsky \& Tortora }{2010}]{NRT10} Napolitano N. R., Romanowsky A. J. \& Tortora,
C. 2010, MNRAS, 405, 2351 (NRT10)

\bibitem[Navarro et al.(1997)]{NFW} Navarro, J.~F., Frenk, C.~S., \& White, S.~D.~M.\ 1997, \apj, 490, 493


\bibitem[Romanowsky et al.(2009)]{Romanowsky+09} Romanowsky, A.~J., Strader, J., Spitler, L.~R., Johnson, R., Brodie, J.~P., Forbes, D.~A., \& Ponman, T.\ 2009, AJ, 137, 4956



\bibitem[\protect\citeauthoryear{Salpeter}{1955}]{Salpeter55} Salpeter, E.E. 1955 ApJ, 121, 161

\bibitem[\protect\citeauthoryear{Spergel et al.}{2007}]{WMAP2} Spergel, D.~N., et al. 2007, ApJS, 170, 377


\bibitem[\protect\citeauthoryear{Thomas et al.}{2005}]{Thomas2005} Thomas, J., Saglia, R. P., Bender, R., Thomas, D., Gebhardt, K.,
Magorrian, J., Corsini, E. M., Wegner, G. 2005, MNRAS, 360, 1355


\bibitem[\protect\citeauthoryear{Thomas et al.}{2009}]{Thomas+09} Thomas, J., Saglia, R. P., Bender, R., Thomas, D., Gebhardt, K.,
Magorrian, J., Corsini, E. M. \& Wegner, G. 2009, ApJ, 691, 770


\bibitem[\protect\citeauthoryear{Tortora et al.}{2007}]{Tortora2007} Tortora C., Cardone V.F. \& Piedipalumbo E. 2007, A\&A, 463, 105


\bibitem[\protect\citeauthoryear{Tortora et al.}{2009}]{Tortora2009a} Tortora C. et al. 2009, MNRAS, 396,
1132 (T+09)

\bibitem[\protect\citeauthoryear{Treu et al.}{2010}]{Treu+09} Treu, T., Auger, M. W., Koopmans, L. V. E., Gavazzi, R., Marshall,
P. J. \& Bolton, A. S. 2010, ApJ, 709, 119


\bibitem[Walker et al.(2010)]{Walker+10b} Walker, M.~G., McGaugh, S.~S., Mateo, M., Olszewski, E., \& Kuzio de Naray, R.\ 2010, ApJL, in press, arXiv:1004.5228




\end{thebibliography}
\end{document}